\documentclass[a4paper]{article}
\usepackage[margin=1in]{geometry}
\usepackage{microtype}
\usepackage{amsmath,amssymb}
\usepackage{graphicx}
\usepackage{wrapfig}
\usepackage{array}
\usepackage[algoruled,linesnumbered,algo2e,vlined]{algorithm2e}
\usepackage{amssymb}
\usepackage{url}
\usepackage{multirow}
\usepackage{cite}

\newtheorem{theorem}{Theorem}

\newtheorem{lemma}[theorem]{Lemma}

\newcommand{\emptystr}{\varepsilon}

\newcommand{\BWT}{\mathsf{BWT}}
\newcommand{\RLBWT}{\mathsf{RLBWT}}
\newcommand{\rank}{\mathsf{rank}}
\newcommand{\select}{\mathsf{select}}
\newcommand{\access}{\mathsf{access}}
\newcommand{\occ}{\mathsf{occ}}
\newcommand{\ins}{\mathsf{insert}}
\newcommand{\del}{\mathsf{delete}}

\newcommand{\psds}{\mathsf{T}}
\newcommand{\pssearch}{\mathsf{search}}
\newcommand{\pssum}{\mathsf{sum}}
\newcommand{\psupdate}{\mathsf{update}}
\newcommand{\sizeA}{\mathit{LA}}
\newcommand{\psumA}{\mathit{WA}}

\newcommand{\wt}{\mathsf{H}}
\newcommand{\xtxt}{\mathit{X}}
\newcommand{\psdsall}{\psds_{\mathit{all}}}

\title{
  A Faster Implementation of Online Run-Length Burrows-Wheeler Transform
}
\author{
   Tatsuya Ohno
   \and
   Yoshimasa Takabatake
   \and
   Tomohiro I
   \and
   Hiroshi Sakamoto\\
   {Kyushu Institute of Technology, Japan}\\
   {\texttt{\{t\_ohno, takabatake\}@donald.ai.kyutech.ac.jp}},\\
   {\texttt{\{tomohiro, hiroshi\}@ai.kyutech.ac.jp}}\\
}

\date{}
\begin{document}
\maketitle

\begin{abstract}
Run-length encoding Burrows-Wheeler Transformed strings, resulting in \emph{Run-Length BWT (RLBWT)},
is a powerful tool for processing highly repetitive strings.
We propose a new algorithm for online RLBWT working in run-compressed space, 
which runs in $O(n\lg r)$ time and $O(r\lg n)$ bits of space, where $n$ is the length of input string $S$ received so far
and $r$ is the number of runs in the BWT of the reversed $S$.
We improve the state-of-the-art algorithm for online RLBWT in terms of empirical construction time.
Adopting the dynamic list for maintaining a total order,
we can replace rank queries in a dynamic wavelet tree on a run-length compressed string by 
the direct comparison of labels in a dynamic list.
The empirical result for various benchmarks show the efficiency of our algorithm, especially for highly repetitive strings. 
\end{abstract}

\section{Introduction}\label{sec:intro}
\subsection{Motivation}
The \emph{Burrows-Wheeler Transform (BWT)}~\cite{Burrows1994BWT} is 
one of the most successful and elegant technique for lossless compression.
When a string contains several frequent substrings, the transformed string would have several \emph{run}s, 
i.e., maximal repeat of a symbol.
Then, such a BWT string is easily compressed by run-length compression.
We refer to the run-length compressed string as the \emph{Run-Length BWT (RLBWT)} of the original string.
Because of the definition of BWT, the number $r$ of runs in the RLBWT is closely related to 
the easiness of compression of the original string.
In fact, $r$ can be (up to) exponentially smaller than the text length, and
several studies~\cite{Belazzougui2015CompositeRepAwareDSs,Siren2012PhDthesis,Sir'en2008RCI,MakinenNSV10} showed that $r$ is available for a measure of repetitiveness.

After the invention of BWT, various applications have been proposed for string processing~\cite{FerraginaM00,FerraginaXBWT05,Bowe12}.
The most notable one would be the BWT based self-index, called FM index~\cite{FerraginaM00},
which allows us to search patterns efficiently while storing text in the entropy-based compressed space.
However, the traditional entropy-based compression is not enough to process highly repetitive strings
because it does not capture the compressibility in terms of repetitiveness.
Therefore several authors have studied ``repetitive-aware'' self-indexes based on 
RLBWT~\cite{Belazzougui2015CompositeRepAwareDSs,Siren2012PhDthesis,Sir'en2008RCI,MakinenNSV10}.
In particular, a self-index in~\cite{Belazzougui2015CompositeRepAwareDSs} works in space proportional to the sizes of the RLBWT and LZ77~\cite{LZ77},
another powerful compressor that can capture repetitiveness.

When it comes to constructing the RLBWT,
a major concern is to reduce the working space depending on the repetitiveness of a given text.
Namely, the problem is to construct the RLBWT \emph{online in run-length compressed space}.
It has been suggested in~\cite{MakinenNSV10} that 
we can solve the problem using a dynamic data structure supporting rank queries on run-length encoded strings.
An implementation appears very recently in~\cite{Prezza2017Dynamic,DYNAMIC}, proving its merit in space reduction.
However the throughput is considerably sacrificed probably due to its use of dynamic succinct data structure.
To ameliorate the time consumption, we present a novel algorithm for online RLBWT and 
show experimentally that our implementation runs faster with reasonable increase of memory consumption.
Since Policriti and Prezza~\cite{Policriti2016RLBWTtoLZ77} recently proposed 
algorithms to compute LZ77 factorization in compressed space via RLBWT,
online RLBWT becomes more and more important, and therefore,
practical time-space tradeoffs are worth exploring.

\subsection{Our Contribution}
Given an input string $S = S[1]S[2] \cdots S[n]$ of length $n$ in online manner,
the algorithm described in~\cite{Policriti2016RLBWTtoLZ77} constructs the RLBWT of the reversed string $S^R = S[n]\cdots S[2]S[1]$
in $O(r\lg n)$ bits of space and $O(n\lg r)$ time, where $r$ is the number of
runs appearing in the BWT of $S^R$.
When a new input symbol $c$ is appended, whereas the BWT of $Sc$ requires (in the worst case) sorting all the suffixes again,
the BWT of $(Sc)^R$ requires just inserting $c$ into the BWT of $S^R$,
and the insert position can be efficiently computed by rank operations on the BWT of $S^R$.
Hence a dynamic data structure on a run-length compressed string supporting rank operations allows to construct the RLBWT online.
However, the algorithm of~\cite{Policriti2016RLBWTtoLZ77} internally uses rank operations on dynamic wavelet trees, which is considerably slow in practice.

In order to get a faster implementation, we replace the work carried out on dynamic wavelet trees 
by a comparison of integers using the dynamic maintenance of a total order.
Here, the \emph{Order-Maintenance Problem} is to maintain a total order of elements
subject to $\mathit{insert}(X,Y)$: insert a new element $Y$ immediately after $X$ in the total order,
$\mathit{delete}(X)$: remove $X$ from the total order, and $\mathit{order}(X,Y)$: determine whether $X>Y$ in the total order.
Bender et al.~\cite{Bender2002TSA} proposed a simple algorithm for this problem to allow $O(1)$ amortized insertion
and deletion time and $O(1)$ worst-case query time.
Adopting this technique, we develop a novel data structure for computing 
the insert position of $c$ in the current BWT by a comparison of integers, instead of heavy rank operations on dynamic wavelet trees.

Compared to the baseline~\cite{Policriti2016RLBWTtoLZ77}, we significantly improve the throughput of RLBWT with reasonable increase of memory consumption.
Although there is a tradeoff between memory consumption and throughput performance,
as shown in the experimental results,
the working space of our algorithm is still sufficiently smaller than the input size, especially for highly repetitive strings.

\section{Preliminaries}\label{sec:prelim}
Let $\Sigma$ be an ordered \emph{alphabet}.
An element of $\Sigma^*$ is called a \emph{string}.
The length of a string $S$ is denoted by $|S|$. 
The empty string $\emptystr$ is the string of length 0,
namely, $|\emptystr| = 0$.
For a string $S = XYZ$, strings $X$, $Y$, and $Z$ are called 
a \emph{prefix}, \emph{substring}, and \emph{suffix} of $S$,
respectively.
For $1 \leq i \leq |S|$, the $i$th character of a string $S$ is denoted by $S[i]$.
For $1 \leq i \leq j \leq |S|$, 
let $S[i..j] = S[i] \cdots S[j]$, i.e., $S[i..j]$ is the substring of $S$
starting at position $i$ and ending at position $j$ in $S$. 
For convenience, let $S[i..j] = \emptystr$ if $j < i$.

In the \emph{run-length encoding (RLE)} of a string $S$, 
a maximal run $c^e$ (for some $c \in \Sigma$ and $e \in \mathcal{N}$) of a single character in $S$ is encoded by a pair $(c, e)$,
where we refer to $c$ and respectively $e$ as the \emph{head} and \emph{exponent} of the run.
Since each run is encoded in $O(1)$ words (under Word RAM model with word size $\Omega(\lg |S|)$), 
we refer to the number of runs as the size of the RLE.\@
For example, $S = \mathtt{aaaabbcccacc} = \mathtt{a}^{4}\mathtt{b}^2\mathtt{c}^3\mathtt{a}^1\mathtt{c}^2$
is encoded as $(\mathtt{a}, 4), (\mathtt{b}, 2), (\mathtt{c}, 3), (\mathtt{a}, 1), (\mathtt{c}, 2)$, and the size of the RLE is five.

For any string $S$ and any $c \in \Sigma$, let $\occ_{c}(S)$ denote the number of occurrences of $c$ in $S$.
Also, let $\occ_{<c}(S)$ denote the number of occurrences of any character smaller than $c$ in $S$, i.e., $\occ_{<c}(S) = \sum_{c' < c} \occ_{c'}(S)$.
For any $c \in \Sigma$ and position $i~(1 \leq i \leq |S|)$, 
$\rank_{c}(S, i)$ denotes the number of occurrences of $c$ in $S[1..i]$, i.e., $\rank_{c}(S, i) = \occ_{c}(S[1..i])$.
For any $c \in \Sigma$ and $i~(1 \leq i \leq \occ_{c}(S))$, 
$\select_{c}(S, i)$ denotes the position of the $i$th $c$ in $S$, i.e., $\select_{c}(S, i) = \min \{ j \mid \rank_{c}(S, j) = i \}$.
Also we let $\access(S, i)$ denote the query to ask for $S[i]$.
We will consider data structures to answer $\occ_{<c}$, $\rank$, $\select$, and $\access$ without having $S$ explicitly.

\subsection{BWT}\label{sec:bwt}
Here we define the BWT of a string $S \in \Sigma^+$, denoted by $\BWT_{S}$.
For convenience, we assume that $S$ ends with a terminator $\$ \in \Sigma$ 
whose lexicographic order is smaller than any character in $S[1..|S|-1]$.
$\BWT_{S}$ is obtained by sorting all non-empty suffixes of $S$ lexicographically and 
putting the immediately preceding character of each suffix (or $\$$ if there is no preceding character) in the order.

For the online construction of BWT, it is convenient to consider ``prepending'' (rather than appending) a character to $S$ 
because it does not change the lexicographic order among existing suffixes.\footnote{Or appending a character but constructing BWT for reversed string.}
Namely, for some $c \in \Sigma$, we consider updating $\BWT_{S}$ to $\BWT_{cS}$ efficiently.
The task is to replace the unique occurrence of $\$$ in $\BWT_{S}$ with $c$, and insert $\$$ into appropriate position.
Since replacing can be easily done if we keep track of the current position of $\$ $,
the main task is to find the new position of $\$$ to insert, which can be done with a standard operation on BWT as follows:
Let $i$ be the position of $\$$ in $\BWT_{S}$, then the new position is computed by $\rank_{c}(\BWT_{S}, i) + \occ_{<c}(S) + 1$
because the new suffix $cS$ is the $(\rank_{c}(\BWT_{S}, i) + 1)$th lexicographically smallest suffix among those starting with $c$,
and there are $\occ_{<c}(S)$ suffixes starting with some $c'~(< c)$.
Thus, BWT can be constructed online using a data structure that supports rank, $\occ_{<c}$, and insert queries.

Let $\RLBWT_{S}$ denote the run-length encoding of $\BWT_{S}$.
In Section~\ref{sec:algo}, 
we study data structures that supports $\rank_{c}$, $\occ_{<c}$ and insert queries on run-length encoded strings,
which can be directly used to construct $\RLBWT_{S}$ online in $O(|S| \lg r)$ time and $O(r \lg |S|)$ bits of space, 
where $r$ is the size of RLE of $\BWT_{S}$.

\subsection{Searchable partial sums with indels}
We use a data structure for the \emph{searchable partial sums with indels (SPSI)} problem as a tool.
The SPSI data structure $\psds$ ought to maintain a dynamic sequence 
$Z[1..m]$ of non-negative integers (called \emph{weights}) to support the following queries as well as insertion/deletion of weights:
\begin{itemize}
\item $\psds.\pssum(k)$: Return the partial sum $\sum_{j = 1}^{k}Z[j]$.
\item $\psds.\pssearch(i)$: For an integer $i~(1 \leq i \leq \psds.\pssum(m))$, 
      return the minimum index $k$ such that $\psds.\pssum(k) \geq i$.
\item $\psds.\psupdate(k, \delta)$: For a (possibly negative) integer $\delta$ with $Z[k] + \delta \geq 0$, update $Z[k]$ to $Z[k] + \delta$.
\end{itemize}

We employ a simple implementation of $\psds$ based on a B+tree whose $k$th leaf corresponds to $Z[k]$.\footnote{More sophisticated solutions can be found in~\cite{Hon2011SuccinctDS_SearchablePartialSums,Navarro2014FFSa,Bille2016DynamicRelativeCompress_DynPartialSum_SubstrConcat}, but none of them has been implemented to the best of our knowledge.}
Let $B~(\geq 3)$ be the parameter of B+trees that represents the arity of an internal node.
Namely the number of children of each internal node ranges from $B/2$ to $B$ (unless $m$ is too small),
and thus, the height of the tree is $O(\log_{B} m)$.
An internal node has two integer arrays $\sizeA$ and $\psumA$ of length $B$ such that
$\sizeA[j]$ (resp.\ $\psumA[j]$) stores the sum of \#leaves (resp.\ weights) under the subtrees of up to the $j$th child of the node.

Using these arrays, we can easily answer $\psds.\pssum$ and $\psds.\pssearch$ queries in $O(\log_{B} m)$ time 
while traversing the tree from the root to a leaf:
For example, $\psds.\pssum(k)$ can be computed by traversing to the $k$th leaf (navigated by $\sizeA$)
while summing up the weights of the subtrees existing to the left of the traversed path by $\psumA$.
It is the same for $\psds.\pssearch(i)$ (except switching the roles of $\sizeA$ and $\psumA$).
For $\psds.\psupdate(k, \delta)$ query, 
we only have to update $\sizeA$ and $\psumA$ of the nodes in the path from the root to the $k$th leaf, which takes $O(B \log_{B} m)$ time.
Also, indels can be done in $O(B \log_{B} m)$ time with standard split/merge operations of B+trees.

Naively the space usage is $O(m \lg M)$ bits, where $M$ is the sum of all weights.
Here we consider improving this to $O(m \lg (M/m))$ bits.
Let us call an internal node whose children are leaves a \emph{bottom node},
for which we introduce new arity parameter $B_{L}$, differentiated from $B$ for the other internal nodes.
For a bottom node, we discard $\sizeA$, $\psumA$ and the pointers to the leaves.
Instead we let it store the weights of its children in a space efficient way.
For example, using gamma encoding, the total space usage for the bottom nodes becomes $O(\sum_{j = 1}^{m} \lg Z[j]) = O(m \lg (M/m))$ bits.
The other (upper) part of $\psds$ uses $O(m \lg M / B_{L})$ bits, which can be controlled by $B_{L}$.
The queries can be supported in $O(B_{L} + B \log_{B} m / B_{L})$ time.
Hence, setting $B = O(1)$ and $B_{L} = \Theta(\lg m)$, we get the next lemma.
\begin{lemma}\label{lemma:spsi}
  For a dynamic sequence of weights,
  there is a SPSI data structure of $O(m \lg (M/m))$ bits supporting queries in $O(\lg m)$ time,
  where $m$ is the current length of the sequence and $M$ is the sum of weights.
\end{lemma}

\section{Dynamic Rank/Select Data Structures on Run-length Encoded Strings}\label{sec:algo}
In this section, we study dynamic rank/select data structures working on run-length encoded strings.
Note that $\select$ and delete queries are not needed for online RLBWT algorithms, but we also provide them as they may find other applications.
Throughout this section, we let $\xtxt$ denote the current string with $n = |\xtxt|$, RLE size $r$, and containing $\sigma$ distinct characters.
We consider the following update queries as well as
$\rank_{c}$, $\select_{c}$, $\access$ and $\occ_{<c}$ queries on $\xtxt$:
\begin{itemize}
  \item $\ins(\xtxt, i, c^e)$: For a position $i~(1 \leq i \leq n+1)$, $c \in \Sigma$ and $e \in \mathcal{N}$, 
        insert $c^e$ between $\xtxt[i-1]$ and $\xtxt[i]$, i.e., $\xtxt \leftarrow \xtxt[1..i-1] c^e \xtxt[i..n]$.
  \item $\del(\xtxt, i, e)$: For a position $i~(1 \leq i \leq n-e+1)$ such that $\xtxt[i..i+e-1] \in c^e$ for some $c \in \Sigma$,
        delete $\xtxt[i..i+e-1]$, i.e., $\xtxt \leftarrow \xtxt[1..i-1] \xtxt[i+e..n]$.
\end{itemize}

\begin{theorem}\label{theorem:ds}
  There is a data structure that occupies $O(r \lg n)$ bits of space 
  and supports $\rank_{c}$, $\select_{c}$, $\access$, $\occ_{<c}$, $\ins$ and $\del$ in $O(\lg r)$ time.
\end{theorem}

We will describe two data structures holding the complexities of Theorem~\ref{theorem:ds}
in theory but likely exhibiting different time-space tradeoffs in practice.
In Subsection~\ref{sec:algo_pp}, we show an existing data structure.
On the basis of this data structure, in Subsection~\ref{sec:algo_new},
we present our new data structure to get a faster implementation.

We note that the problem to support $\occ_{<c}$ in $O(\sigma \lg n)$ bits of space and $O(\lg \sigma)$ time is somewhat standard.
For instance, we can think about the SPSI data structure of Lemma~\ref{lemma:spsi} storing $\occ_{c}(\xtxt)$'s in increasing order of $c$.
It is easy to modify the data structure so that, for a given $c$,
we can traverse from the root to the leaf corresponding to the predecessor of $c$,
where we mean by the predecessor of $c$ the largest character $c'$ that is smaller than $c$ and appears in $\xtxt$.
Then $\occ_{<c}$ queries can be supported in a similar way to $\pssum$ queries using $\psumA$.
Thus in the following subsections, we focus on the other queries.

\subsection{Existing data structure}\label{sec:algo_pp}
Here we review the data structure described in~\cite{Policriti2016RLBWTtoLZ77} with implementation available in~\cite{Prezza2017Dynamic,DYNAMIC}.\footnote{The basic idea of the algorithm originates from the work of RLFM+ index in~\cite{MakinenNSV10}.}
In theory it satisfies Theorem~\ref{theorem:ds}
though its actual implementation has the time complexity of $O(\lg \sigma \lg r)$ slower than $O(\lg r)$.

\begin{figure}[t]
  \begin{center}
  \includegraphics[scale=0.315]{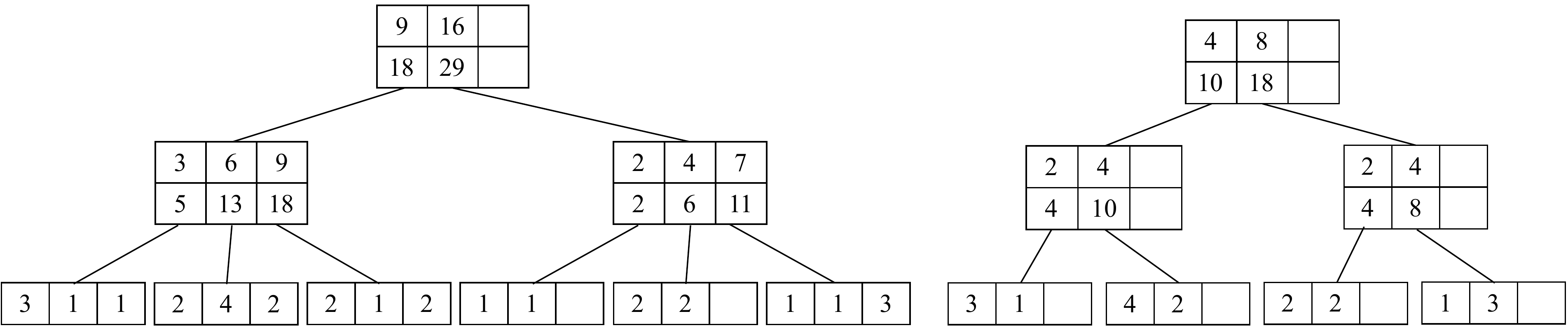}
  \end{center}
  \caption{
    For $\xtxt = \mathtt{a}^{3}\mathtt{b}^{1}\mathtt{a}^{1}\mathtt{c}^{2}\mathtt{a}^{4}\mathtt{b}^{2}\mathtt{a}^{2}\mathtt{c}^{1}\mathtt{a}^{2}\mathtt{b}^{1}\mathtt{c}^{1}\mathtt{a}^{2}\mathtt{c}^{2}\mathtt{a}^{1}\mathtt{b}^{1}\mathtt{a}^{3}$,
    examples of $\psdsall$ (left) and $\psds_{\mathtt{a}}$ (right) with $B = B_{L} = 3$ are shown.
    Note that the other components of the data structure ($\psds_{\mathtt{b}}$, $\psds_{\mathtt{c}}$ and $\wt$) are omitted here.
    $\psdsall$ holds the sequence $[3, 1, 1, 2, 4, 2, 2, 1, 2, 1, 1, 2, 2, 1, 1, 3]$ of the exponents in its leaves,
    and $\psds_{\mathtt{a}}$ holds the sequence $[3, 1, 4, 2, 2, 2, 1, 3]$ of the exponents of $a$'s runs in its leaves.
    For a node having two rows, the first row represents $\sizeA$ and the second $\psumA$.
  }
\label{fig:PP}
\end{figure}

Let $(c_1, e_1), (c_2, e_2), \ldots, (c_r, e_r)$ be the RLE of $\xtxt$.
The data structure consists of three components (see also Fig.~\ref{fig:PP} for the first two):
\begin{enumerate}
\item $\psdsall$: SPSI data structure for the sequence $e_1 e_2 \cdots e_r$ of all exponents.
\item $\psds_{c}$ (for every $c \in \Sigma$): SPSI data structure for the sequence of the exponents of $c$'s run.
\item $\wt$: Dynamic rank/select data structure for the head string $H = c_1 c_2 \cdots c_r$.
      There is a data structure (e.g., see~\cite{Navarro2014ODS,Munro2015CompressedDS_DynamicSeq}),
      with which $\wt$ can be implemented in $r \lg \sigma + o(r \lg \sigma) + O(\sigma \lg r)$ bits while supporting queries in $O(\lg r)$ time.
      (However, the actual implementation of~\cite{Prezza2017Dynamic,DYNAMIC} employs a simpler algorithm based on wavelet trees that has $O(\lg \sigma \lg r)$ query time.)
\end{enumerate}
Note that for every run $c^e$ there are two copies of its exponent, one in $\psdsall$ and the other in $\psds_{c}$.
Since $\sigma \leq r \leq n$ holds, the data structure (excluding $\occ_{<c}$ data structure) uses 
$r \lg \sigma + o(r \lg \sigma) + O(r \lg (n/r) + \sigma \lg r) = O(r \lg n)$ bits.

Let us demonstrate how to support $\rank_{c}(\xtxt, i)$.
Firstly by computing $k \leftarrow \psdsall.\pssearch(i)$ we can find that $\xtxt[i]$ is in the $k$th run.
Next by computing $k_{c} \leftarrow \wt.\rank_{c}(H, k)$ we notice that, up to the $k$th run, there are $k_{c}$ runs with head $c$.
Here we can check if the head of the $k$th run is $c$, and compute the number $e$ of $c$'s in the $k$th run appearing after $\xtxt[i]$.
Finally, $\psds_{c}.\pssum(k_{c}) - e$ tells the answer of $\rank_{c}(\xtxt, i)$.
It is easy to see that each step can be done in $O(\lg r)$ time.

Note that $\wt$ plays an important role to bridge two trees $\psdsall$ and $\psds_{c}$ by converting the indexes $k$ and $k_{c}$.
The update queries also use this mechanism:
We first locate the update position in $\psdsall$, then find the update position in $\psds_{c}$ by bridging two trees with $\wt$.
After locating the positions, the updates can be done in each dynamic data structure.
$\select_{c}(\xtxt, i)$ can be answered by first locating $i$th $c$ in $\psds_{c}$, 
finding the corresponding position in $\psdsall$ with $\wt.\select_{c}$,
then computing the partial sum up to the position in $\psdsall$.
Finally, $\access(\xtxt, i)$ is answered by $\wt.\access(H, \psdsall.\pssearch(i))$.

\subsection{New data structure}\label{sec:algo_new}
Now we present our new data structure satisfying Theorem~\ref{theorem:ds}.
We share some of the basic concepts with the data structure described in Section~\ref{sec:algo_pp}.
For example, our data structure also uses the idea of answering queries by going back and forth between $\psdsall$ and $\psds_{c}$.
However we do not use $\wt$ to bridge the two trees.
Succinct data structures (like $\wt$) are attractive if the space limitation is critical,
but otherwise the suffered slow-down might be intolerable.
Therefore we design a fast algorithm that bridges the two trees in a more direct manner,
while taking care not to blow up the space too much.

\begin{figure}[t]
  \begin{center}
  \includegraphics[scale=0.40]{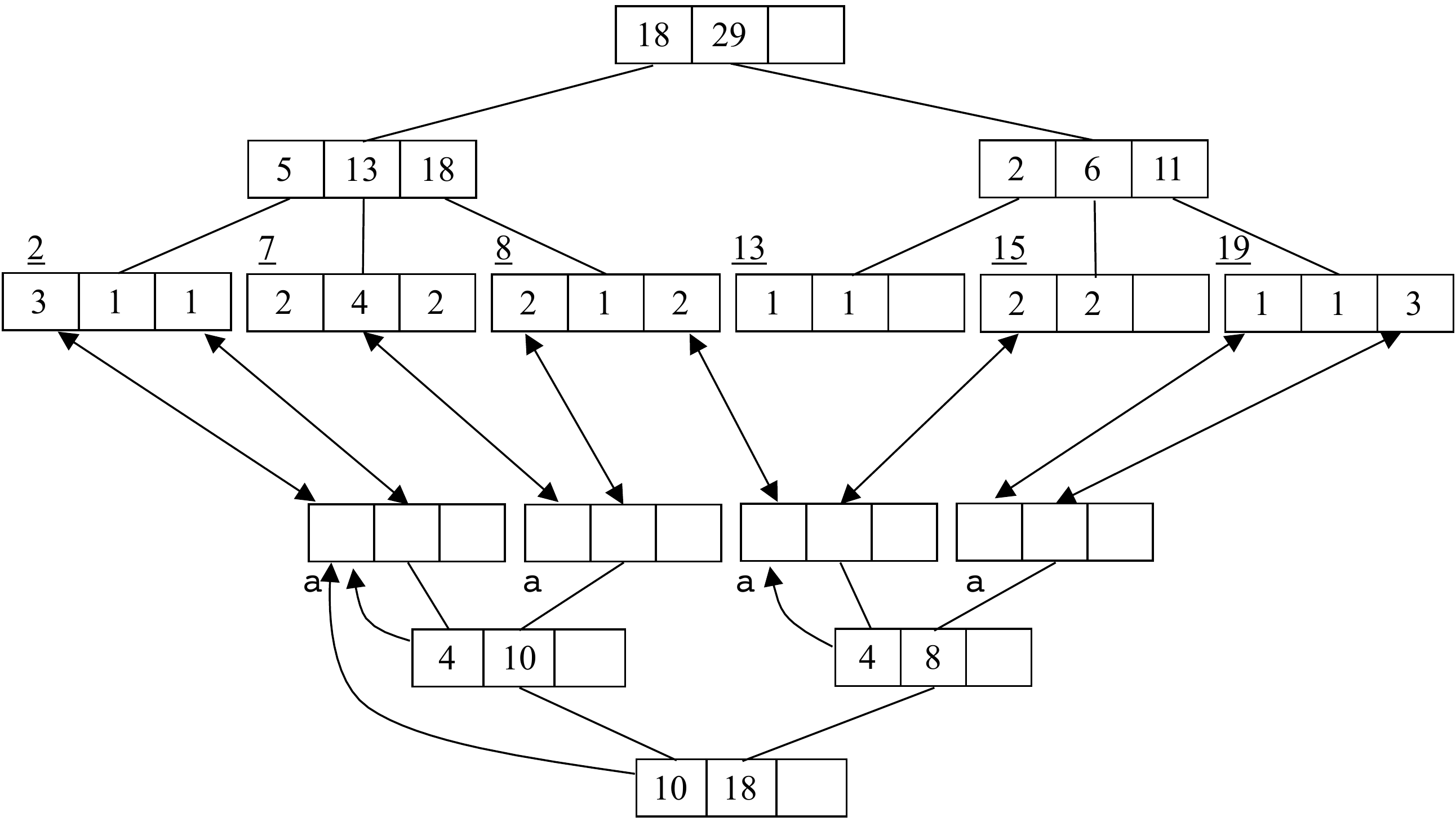}
  \end{center}
  \caption{
    For $X = \mathtt{a}^{3}\mathtt{b}^{1}\mathtt{a}^{1}\mathtt{c}^{2}\mathtt{a}^{4}\mathtt{b}^{2}\mathtt{a}^{2}\mathtt{c}^{1}\mathtt{a}^{2}\mathtt{b}^{1}\mathtt{c}^{1}\mathtt{a}^{2}\mathtt{c}^{2}\mathtt{a}^{1}\mathtt{b}^{1}\mathtt{a}^{3}$ (same as the one in Fig.~\ref{fig:PP}), 
    examples of modified $\psdsall$ (up) and $\psds_{\mathtt{a}}$ (down) with $B = B_{L} = 3$ are shown,
    where $\psds_{\mathtt{a}}$ is illustrated upside down.
    Note that the data structure related to $\psds_{\mathtt{b}}$ and $\psds_{\mathtt{c}}$ (e.g., pointers of Change\ref{change_links} to them) are omitted here.
    Each pair of leaves corresponding to the same run is connected by bidirectional pointers (Change\ref{change_links}).
    Each internal node of $\psds_{\mathtt{a}}$ has pointer to its leftmost leaf (Change\ref{change_ptr}).
    The character $\mathtt{a}$ is stored in each bottom node of $\psds_{\mathtt{a}}$ (Change\ref{change_char}).
    Each bottom node of $\psdsall$ stores a label (underlined number) that is monotonically increasing from left to right (Change\ref{change_label}).
    $\sizeA$s and weights in the leaves of $\psds_{\mathtt{a}}$ are discarded (Change\ref{change_minor}).
  }
\label{fig:ours}
\end{figure}

In order to do without $\wt$, we make some changes to $\psdsall$ and $\psds_{c}$ (see also Fig.~\ref{fig:ours}):
\begin{enumerate}
\item We maintain bidirectional pointers connecting every pair of leaves representing the same run in $\psdsall$ and $\psds_{c}$
      (recall that every run with head $c$ has exactly one corresponding leaf in each of $\psdsall$ and $\psds_{c}$).\label{change_links}
\item For every internal node of $\psds_{c}$, we store a pointer to the leftmost leaf in the subtree rooted at the node.\label{change_ptr}
\item For every bottom node of $\psds_{c}$, we store the character $c$.\label{change_char}
\item For every bottom node of $\psdsall$, we store a label (a positive integer) such that the labels of bottom nodes are monotonically increasing from left to right.
      Since bottom nodes are inserted/deleted dynamically, 
      we utilize the algorithm~\cite{Bender2002TSA} for the order-maintenance problem to maintain the labels.\label{change_label}
\item Minor implementation notes: Every $\sizeA$ can be discarded as our data structure does not use the navigation of indexes.
      Also, we can quit storing the leaf-level weights in $\psds_{c}$ as it can be retrieved using the pointer to the corresponding leaf in $\psdsall$.\label{change_minor}
\end{enumerate}

\subsubsection{Space analysis.}
In the change list, Changes\ref{change_links}-\ref{change_label} increase the space usage while Change\ref{change_minor} reduces.
It is easy to see that the increase fits in $O(r \lg n)$ bits.
More precisely, since Changes\ref{change_ptr}-\ref{change_label} are made to internal nodes,
the increase by these changes is $O(r \lg n / B_{L})$ bits, which is more or less controllable by $B_{L}$
(recall that $B_{L}$ is arity parameter for bottom nodes, 
and we have $O(r \lg n / B_{L}) = O(r \lg (n/r))$ by setting $B_{L} = \Theta(\lg r)$ for Lemma~\ref{lemma:spsi}).
On the other hand, Change\ref{change_links} is made to leaves and takes $2 r \lg r$ bits of space.
Thus, the total space usage of the data structure (excluding $\occ_{<c}$ data structure) is $2 r \lg r + O(r \lg (n/r)) = O(r \lg n)$ bits.

By this analysis, it is expected that $2 r \lg r$ becomes a leading term when the ratio $n/r$ is small, i.e., compressibility in terms of RLE is not high.
It should be compared to $r \lg \sigma + o(r \lg \sigma) + O(r \lg (n/r) + \sigma \lg r)$ bits used by the data structure of Section~\ref{sec:algo_pp},
in which $r \lg r$ term does not exist. Hence, the smaller the ratio $n/r$, the larger the gap between the two data structures in space usage will be.
On the other hand, when the $r \lg (n/r)$ term is leading, i.e., $r$ is sufficiently smaller than $n$, the increase by the $r \lg r$ term would be relatively moderate.

\subsubsection{Answering queries.}
We show how to answer queries on our data structure.
All queries are supported in $O(\lg r)$ time.

$\access(\xtxt, i)$:
We first traverse from the root of $\psdsall$ to the run containing $\xtxt[i]$ (navigated by $\psumA$), 
jump to the corresponding leaf of $\psds_{c}$ by pointer of Change\ref{change_links},
then read the character stored in the bottom node of $\psds_{c}$ due to Change\ref{change_char}.

$\select_{c}(\xtxt, i)$:
We first traverse from the root of $\psds_{c}$ to the run containing $i$th $c$ (navigated by $\psumA$).
At the same time, we can compute the rank $i'$ of $i$th $c$ within the run.
Next we jump to the corresponding leaf in $\psdsall$ by pointer of Change\ref{change_links},
then compute the sum of characters appearing strictly before the leaf while going up the tree.
The answer to $\select_{c}(\xtxt, i)$ is the sum plus $i'$.

$\rank_{c}(\xtxt, i)$:
Recalling the essence of the algorithm described in Section~\ref{sec:algo_pp},
we can answer $\rank_{c}(\xtxt, i)$ if we locate the leaf of $\psds_{c}$ 
representing the rightmost $c$'s run that starts at or before position $i$.
In order to locate such leaf $v$, 
we first traverse from the root of $\psdsall$ to the run containing $\xtxt[i]$ (navigated by $\psumA$).
If we are lucky, we may find a $c$'s run in the bottom node containing $\xtxt[i]$, in which case
we can easily get $v$ or the successor of $v$
by using the pointer of Change\ref{change_links} outgoing from the $c$'s run.
Otherwise, we search for $v$ traversing $\psds_{c}$ from the root navigated by labels of Change\ref{change_label}.
Let $t$ be the label of the bottom node containing $\xtxt[i]$.
Then, it holds that $v$ is the rightmost leaf pointing to a node of $\psdsall$ with label smaller than $t$.
Since the order of labels is maintained, we can use $t$ as a key for binary search, i.e.,
we notice that an internal node $u$ (and its succeeding siblings) cannot contain $v$ 
if the leftmost leaf in the subtree rooted at $u$ points to a node of $\psdsall$ with label greater than $t$.
Using the pointer of Change\ref{change_ptr} to jump to the leftmost leaf,
we can conduct each comparison in $O(1)$ time, and thus, we can find $v$ in $O(\lg r)$ time.

Update queries:
The main task is to locate the update positions both in $\psdsall$ and $\psds_{c}$,
and this is exactly what we did in $\rank_{c}$ query---locating the run containing $\xtxt[i]$ and $v$.
After locating the update positions, the update can be done in $O(\lg r)$ time in each tree.
When the update operation invokes insertion/deletion of a bottom node of $\psdsall$,
we maintain labels of Change\ref{change_label} using the algorithm of~\cite{Bender2002TSA}.
We note that the algorithm of~\cite{Bender2002TSA} takes $O(\lg r)$ amortized time per ``indel of bottom node'', 
and hence, takes $O(1)$ amortized time per ``indel of leaf''
(recall that $B_{L} = \Theta(\lg r)$, and one indel of bottom node needs $\Theta(\lg r)$ indels of leaves).
In addition, the algorithm is quite simple and efficiently implementable without any data structure than labels themselves.

\section{Experiments}

\begin{table}[t]
\caption{Computation time in seconds and working space in mega bytes 
to construct the RLBWT of $r$ runs from each dataset
of size $|S|$ using the proposed method (ours) and the previous method (PP).
}
\label{table:exp}
\begin{center}
\begin{tabular}{|l|r@{.}l|r|rr@{}l|r@{.}l r@{.}l|}
\multirow{2}{*}{dataset}&\multicolumn{2}{c|}{\multirow{2}{*}{$|S|$ (MB)}}&
\multirow{2}{*}{$r$}\ \ \ \ \ \ \ & \multicolumn{3}{c|}{computation time (sec)}
	     & \multicolumn{4}{c|}{working space (MB)}  \\ \cline{5-11}
&\multicolumn{2}{c|}{}&&ours&\multicolumn{2}{c|}{PP}&\multicolumn{2}{r}{ours} &\multicolumn{2}{c|}{PP}\\ \hline \hline
fib41&\ \ 255&503&42&\ \ \ \ \ \ \ \ 27&5&52&0&004&\ \ \ \ \ 0&067\\
rs.13&206&706&76&16&6&23&0&005&0&068\\
tm29&256&000&82&24&8&02&0&005&0&068\\
dblp.xml.00001.1&100&000&172,195&37&2,0&60&2&428&1&307\\
dblp.xml.00001.2&100&000&175,278&37&2,0&70&2&446&1&322\\
dblp.xml.0001.1&100&000&240,376&40&2,1&00&4&381&1&586\\
dblp.xml.0001.2&100&000&269,690&40&2,1&05&4&565&1&730\\
dna.001.1&100&000&1,717,162&58&1,6&67&35&966&5&729\\
english.001.2&100&000&1,436,696&58&2,1&53&20&680&6&166\\
proteins.001.1&100&000&1,278,264&58&1,8&39&19&790&5&133\\
sources.001.2&100&000&1,211,104&49&2,1&41&19&673&5&721\\
cere&439&917&11,575,582&534&7,5&97&186&073&43&341\\
coreutils&195&772&4,732,794&128&4,4&79&81&642&22&301\\
einstein.de.txt&88&461&99,833&30&1,8&07&2&083&1&106\\
einstein.en.txt&445&963&286,697&182&9,2&93&4&836&2&296\\
Escherichia\_Coli&107&469&15,045,277&154&2,0&47&\ \ 316&184&36&655\\
influenza&147&637&3,018,824&91&2,5&01&72&730&12&386\\
kernel&246&011&2,780,095&146&5,3&33&41&758&12&510\\
para&409&380&15,635,177&547&\ \ \ \ \ 7,3&64&329&901&52&005\\
world\_leaders&44&792&583,396&17&8&57&9&335&2&891\\
\hline
boost&1024&000&63,710&320&20,3&27&1&161&0&904\\
samtools&1024&000&562,326&440&21,3&75&9&734&3&595\\
sdsl&1024&000&758,657&419&21,0&14&17&760&4&803\\
\end{tabular}
\end{center}
\end{table}

We implemented in C++ the online RLBWT construction algorithm based on our new rank/select data structure described in Section~\ref{sec:algo_new}
(the source code is available at~\cite{OnlineRLBWT}).
We evaluate the performance of our method comparing with the state-of-the-art implementation~\cite{DYNAMIC}
(we refer to it as PP taking the authors' initials of~\cite{Policriti2016RLBWTtoLZ77})
of the algorithm based on the data structure described in Section~\ref{sec:algo_pp}.
We tested on highly repetitive datasets in repcorpus\footnote{See \url{http://pizzachili.dcc.uchile.cl/repcorpus/statistics.pdf} for statistics of the datasets.}, well-known corpus in this field, and some larger datasets created from git repositories.
For the latter, we use the script~\cite{getgit} to create 1024MB texts
(obtained by concatenating source files from the latest revisions of a given repository, and truncated to be 1024MB)
from the repositories for boost\footnote{\url{https://github.com/boostorg/boost}}, samtools\footnote{\url{https://github.com/samtools/samtools}} 
and sdsl-lite\footnote{\url{https://github.com/simongog/sdsl-lite}} (all accessed at 2017-03-27).
The programs were compiled using g++6.3.0 with -Ofast -march=native option.
The experiments were conducted on a 6core Xeon E5-1650V3 (3.5GHz) machine with 32GB memory running Linux CentOS7.

Table~\ref{table:exp} shows the comparison of the two methods on construction time and working space.
The result shows that our method significantly improves the construction time of PP as we intended.
Especially for dumpfiles of Wikipedia articles (einstein.de.txt and einstein.en.txt), our method ran 60 times faster than PP.\@
Our method also shows good performance for the 1024MB texts from git repositories.
On the other hand, the working space is increased (except the artificial datasets, which are extremely compressible) by 1.3 to 8.7 times.
Especially for less compressible datasets in terms of RLBWT like Escherichia\_Coli, the space usage tends to be worse 
as predicted by space analysis in Section~\ref{sec:algo_new}.
Still for most of the other datasets the working space of our method keeps way below the input size.

\section{Conclusion}

We have proposed an improvement of online construction of RLBWT~\cite{Prezza2017Dynamic,DYNAMIC}, intended to speed up the construction time.
We significantly improved the throughput of original RLBWT with reasonable increase of memory consumption for the benchmarks from various domain.
By applying our new algorithm to the algorithm of computing LZ77 factorization in compressed space using RLBWT~\cite{Policriti2016RLBWTtoLZ77},
we would immediately improve the throughput of~\cite{Policriti2016RLBWTtoLZ77}.
As LZ77 plays a central role in many problems on string processing,
engineering/optimizing implementation for compressed LZ77 computation is important future work.

\section{Acknowledgments}
This work was supported by JST CREST (Grant Number JPMJCR1402), and KAKENHI (Grant Numbers 17H01791 and 16K16009).

\bibliography{ref}

\end{document}